\documentclass[twocolumn, 10pt, superscriptaddress, amsmath, amssymb, longbibliography]{revtex4-1}
\usepackage{graphicx}
\usepackage{changes}
\usepackage{mathtools}
\usepackage{subfig}
\usepackage[pdftex, colorlinks]{hyperref}
\usepackage{color}

\hypersetup{ colorlinks, linkcolor=red, filecolor=red, urlcolor=blue,
  citecolor=blue }

\begin{document}

\title{Energy dissipation in sheared wet granular assemblies}

\author{L. Kovalcinova} \affiliation{Department of Mathematical
  Sciences, New Jersey Institute of Technology, New Jersey, USA}

\author{S. Karmakar} \affiliation{Experimental Physics, Saarland
  University, Saarbr{\"u}cken, D-66123, Germany}

\author{M. Schaber} \affiliation{Experimental Physics, Saarland
  University, Saarbr{\"u}cken, D-66123, Germany}

\author{A.-L. Schuhmacher} \affiliation{Experimental Physics, Saarland
  University, Saarbr{\"u}cken, D-66123, Germany}

\author{M. Scheel} \affiliation{European Synchrotron Radiation
  Facility, Grenoble, F-38043, France} \affiliation{Synchrotron
  Soleil, 91192 Gif-sur-Yvette, France}

\author{M. DiMichiel} \affiliation{European Synchrotron Radiation
  Facility, Grenoble, F-38043, France}

\author{M. Brinkmann} \affiliation{Experimental Physics, Saarland
  University, Saarbr{\"u}cken, D-66123, Germany}

 \author{R. Seemann} \affiliation{Experimental Physics, Saarland
   University, Saarbr{\"u}cken, D-66123, Germany}

\author{L. Kondic} \affiliation{Department of Mathematical Sciences,
  New Jersey Institute of Technology, New Jersey, USA}

\thanks{Corresponding author:}
\email{kondic@njit.edu}
\date{\today}
\begin{abstract}
  Energy dissipation in sheared dry and wet granulates is considered
  in the presence of an externally applied confining pressure.
  Discrete element simulations reveal that for sufficiently small
  confining pressures, the energy dissipation is dominated by the effects
  related to the presence of cohesive forces between the particles.
  The residual resistance against shear can be quantitatively explained by a
  combination of two effects arising in a wet granulate: i) enhanced
  friction at particle contacts in the presence of attractive
  capillary forces, and ii) energy dissipation due to the rupture and
  reformation of liquid bridges. Coulomb friction at grain contacts
  gives rise to an energy dissipation which grows linearly with
  increasing confining pressure, for both dry and wet granulates. Because of
  a lower Coulomb friction coefficient in the case of wet grains, as
  the confining pressure increases the
  energy dissipation for dry systems is faster than for wet ones.
  \end{abstract}

\maketitle

\newpage
\section{Introduction}

The mechanics of wet granulates plays a prominent role in various
fields of process engineering, including the production of
pharmaceutics \cite{Faure2001, Leuenberger2001A, Paul1999, Komlev2002},
wet granulation of powders \cite{Simons2000,Litster2004, Realpe2007},
sintering \cite{Shinagawa1998} and food
production \cite{Bhandari2013}. Owing to this outstanding importance,
a large number of experimental studies and physical models have been
devoted to the mechanics of wet granular matter,
e.g. \cite{Geminard1999, Fournier_05, Mitarai2006,Scheel2008A,
  Fiscina2012, Fall2014}. The transport of stresses in a dry granulate
is governed by an interplay between frictional and repulsive forces acting
between the constituting grains. Dry granulates easily flow under
external forces such as gravity and hardly resist to shear. However, a
confining stress applied to the grains at the surface of the assembly
can reversibly turn a dry granulate into a solid--like material
\cite{Brown2010}. Hence externally applied confining stresses alter
the mechanics of a granular assembly.  A change of the mechanical
properties also occurs when dry grains are mixed with a small amount of a wetting liquid.
Granular assembly then turns into a plastically deformable material, which can sustain finite
tensile and shear stresses \cite{Halsey1998}.

\begin{figure}[ht!]
 \centering
 \includegraphics[width = 0.45\textwidth]{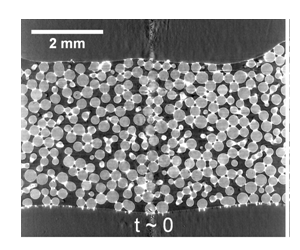}
 \caption{$2$D slice through a $3$D x-ray tomography of a sheared wet glass bead assembly~\cite{Karmakar_unpub},
 see also~\cite{Fournier_05, Herminghaus2005}.
 The gray level indicates the different phases (white: aqueous $\rm ZnI_2$ solution; gray: glass beads; black air).}
\label{fig:exp}
\end{figure}

In this paper, we explore the rheology of dry and wet granulates in
the presence of an externally applied confining stress. To quantify
the resistance to shear as a function of the confining stress, we
determine the energy dissipated in an assembly of particles over a
stationary shear cycle. We perform Discrete Element Simulations (DES) of dense
granular packs with shearing protocol inspired by recent experiments of the 3D pack of wet and dry glass
beads~\cite{Herminghaus2005, Fournier_05}.  Figure~\ref{fig:exp} shows a typical snapshot
obtained using 3D tomography.

From the simulations, we extract the information about the
source of energy dissipation due to direct particle-particle
interaction (friction, inelastic interactions) as well as due to breaking
and reformation of capillary bridges. The analysis of various
contributions to the total energy dissipation provides an insight into the
differences between dry and wet granular sheared packs and the main sources of energy
loss.  The main finding of our DES is that, for small
 applied pressures, the energy dissipated during the process of
 breakage and reformation of the capillary bridges is a main source of energy
 dissipation, dominating both the dissipation due to direct
 particle--particle interaction and the dissipation arising in the
 presence of capillary cohesion. However, for large applied pressure,
 friction dominates the particle--particle interaction for both wet and
 dry granulates. Accordingly, the work to shear a dry granulate becomes
 larger than that to shear a wet one for sufficiently large confining
 pressure.

 This paper is organized as follows.  In Section~\ref{sec:methods}
 we describe the setup of DES and discuss various energy loss mechanisms
 for wet and dry systems.  In Section~\ref{sec:results} we focus on energy
 dissipation mechanisms, discussing in particular the contributions due to
 non-affine motion of the particles and internal cohesion.   We summarize
 the results in Section~\ref{sec:conclusion}.

\section{Methods}
\label{sec:methods}

\subsection{Computational model}

Discrete Element Simulations (DES) of
two--dimensional packs of circular particles that are subject to shear deformations
were carried out with a setup close to the experiments described in~\cite{Fournier_05}.
The aim of this study is to reveal the fundamental mechanisms that control the dissipation in
sheared assemblies of wet and dry particles.
Details of the simulation techniques for dry granular
matter could be found in e.g.~Ref.~\cite{kovalcinova_percolation} and Appendix~\ref{app_sim}.
For later convenience we express all the quantities
used in simulations in terms of the following scales: average particle diameter, $\bar d$, as
the length--scale, average particle mass, $m$, as mass scale, and the
binary particle collision time, $\tau_{\rm c}=\pi\sqrt{ \bar d /2gk_{\rm n}}$, as
the time scale. The parameter $k_{\rm n}$ corresponds to the normal spring
constant between two colliding particles and $g$ is the acceleration
of gravity. The parameters entering the force model can be
connected to physical properties (Young modulus, Poisson ratio) as
described, e.g.~in Ref.~\cite{kondic_99}. The inter--particle friction
coefficient for dry and wet assemblies is $\mu_{\rm dry}=0.29$ and
$\mu_{\rm wet}=0.25$, respectively, to capture different friction properties of the
granular material when a small amount of liquid is added between particles~\cite{Fournier_05}.
Furthermore, we use $k_{\rm n} = 4\times 10^3$, and the coefficient of restitution, as a measure
of the inelasticity of collisions, is $e=0.5$.

The motivation for choosing the value of $k_{\rm n}$ that is
smaller than appropriate for the glass beads used in the
experiments~\cite{Fournier_05} is the computational complexity: the simulations need to be
carried out for long times in physical units, increasing
computational cost; the use of softer particles allows for the use
of larger computational time steps. To confirm that only
quantitative features of the results are influenced by this choice,
we have carried out limited simulations with stiffer particles,
that led to similar results as the ones presented here.

In modeling capillary cohesion, we are
motivated by the experiments~\cite{Karmakar_unpub}, see also Fig.~\ref{fig:exp}, and
employ the capillary force model for
three dimensional (3D) pendular bridges proposed by Willet et
al.~\cite{Willett2000}.   We motivate the choice of the force
model between 3D spheres by the effort to use the same
type of cohesive interaction between particles as the one
expected in experiments.  Note that according to~\cite{Urso1999},
cohesive force in 2D assumes a local maximum at a non-zero
distance of the particles, in contrast to the 3D model.   Furthermore,
to simplify the implementation, we use the approximate expression,
Eq.~(12) in~\cite{Willett2000, Herminghaus2005}, given here in nondimensional form
\begin{equation}
  F_{\rm c,ij}={\frac{\pi d\sigma\cos\theta}
    {1+1.05\hat S_{\rm i,j}+2.5\hat S^2_{\rm i,j}}}\, ,
  \label{eq:cohesive_force}
\end{equation}
where $\hat S_{\rm i,j}=S_{\rm i,j}\sqrt{1/(2V)}$ and
$S_{\rm i,j}=r_{\rm i,j}-(d_{\rm i}+d_{\rm j})/2$ is the
separation of the particle surfaces, where $r_{\rm i,j}$ is
the distance between the centers of the circular particles $i,~j$. The inverse
value of the reduced diameter is
${1 /d}=({1 /d_{\rm i}}+ {1 /d_{\rm j}})/2$ and
$d_{\rm i},~d_{\rm j}$ are the particle diameters. The maximum separation, $S^{\rm max }$, at which a
capillary bridge breaks is given by
\cite{Willett2000}:
\begin{equation}
  S^{\rm max}=(2+{\theta})\left(\frac{V^{ 1/3}}
    {d}+\frac{2V^{ 2/3}}{d^{2}}\right)\, ,
  \label{eq:max_separation}
\end{equation}
where $V$ is the non--dimensional capillary bridge
volume. During a collision we set $S_{\rm i,j}=0.0$ (even when $r_{\rm i,j}<(d_{\rm i}+d_{\rm j})/2$)
since the cohesive force has a constant value when the
particles are in contact (regardless of the amount of compression resulting
from collision)~\cite{Herminghaus2005}.

For the contact angle, $\theta$, and the surface tension,
$\sigma = {\bar \sigma\tau_{\rm c}^2/m}$, we use $\theta=12^\circ$ and
$\bar \sigma=72$~mN/m motivated by the parameters of the experiments
in~\cite{Fournier_05}. The mass is computed from the density of a
glass bead ($\rho=2.5\cdot 10^3$ kg/m$^3$) of average diameter
$\bar d$.

All capillary bridges in expressions (\ref{eq:cohesive_force}) and
(\ref{eq:max_separation}) are assumed to have equal liquid volume
$\bar V=7.4\cdot 10^{-3}\,\bar d^3$.  This value corresponds to the average value in a
3D pack of monodisperse spherical beads with average
diameter $ \bar d$ with a liquid content of $W=2.5\%$~\cite{Herminghaus2005} with respect to the total volume.

A bridge forms after two particles touch and breaks when the bridge
length exceeds the maximum surface--to--surface separation
$S^{\rm max}$. The energy dissipated during a
full cycle of formation and rupture is computed by integrating the
bridge force $F_{\rm c}$ between $S=0$ and $S=S^{\rm max}$, and can be
expressed in closed form as:
\begin{eqnarray}
  E_{\rm c}^{ \rm i,j}&=&4\pi\sigma\cos \theta
                        \,\sqrt\frac{d V}{ 2\epsilon}
                        \Biggl [
                        \arctan\biggl( 5\bar S\sqrt\frac{d}{ 2V\epsilon}
                        + \frac{\delta}{\sqrt{\epsilon}} \biggr) \nonumber\\
                    &-& \arctan\biggl\{ \frac{\delta}{\sqrt{\epsilon}} \biggr\}\Biggr] \label{eq_ec},
\end{eqnarray}
with numerical constants $\delta=1.05$ and $\epsilon=8.8795$.

\subsection{Simulation protocol}

\begin{figure}
  \includegraphics[width=0.45\textwidth]{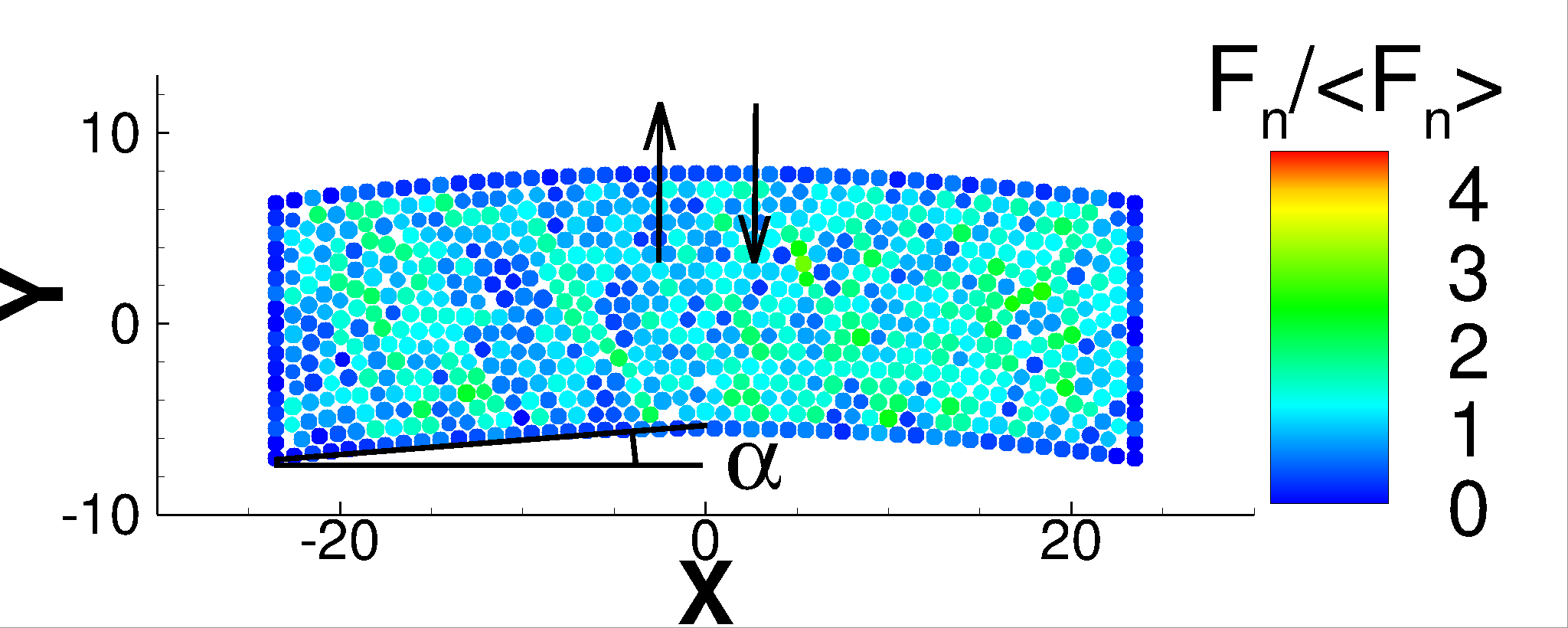}
  \caption{Snapshot of a granular domain with the particles colored according to the
  total normal force normalized by the average normal force imposed on the particles.
  The arrows show the upward and downward direction of the shear.
  For animation, see~\cite{supp_movie}.} \label{chap:parabolic_shear:sim_shear}

\end{figure}
The simulation protocol is set up in such a way that closely follows the
experimental one~\cite{Herminghaus2005, Fournier_05, Karmakar_unpub}.
Figure~\ref{chap:parabolic_shear:sim_shear}
shows the example of a granular system during the shear.
The simulation domain is initially
rectangular and of the size $L_{\rm x},~L_{\rm y}$, with $L_{\rm x} = 47$, and $L_{\rm y} = 17$, both in units of $\bar d$.
The walls are composed of monodisperse particles of size $\bar d$ and mass $m$. The system
particles are chosen randomly from a uniform distribution with mean
$\bar d$ and width $0.4\bar d$. Initially,
the particles are placed on a rectangular grid and initialized with
random velocities. Then, the top and bottom walls are moved inward by
applying initial external pressure, $P_{\rm init}$, until equilibrium is reached and the position of the
top and bottom wall stays fixed. At this point, we start
shearing the system by prescribing a parabolic wall shape evolving in
time.  The maximum value of the shearing angle, $\alpha_m$, defined as the angle between the line connecting the endpoints of the left and right walls and the center of the bottom wall (see Fig.~\ref{chap:parabolic_shear:sim_shear}), is $\alpha_m=4^\circ$.
The motion of the top/bottom wall is periodic in time with period $T$. At the beginning of a cycle, the system is sheared
from the flat state ($\alpha = 0$) in the positive vertical direction. After reaching $\alpha_m$, the shear continues in
the opposite (negative) direction, until $\alpha$ reaches the value $-\alpha_m$ and the direction of the shear is reversed.
The cycle is complete when the system reaches $\alpha = 0$.  More precisely, the
motion of the top/bottom wall over time is given by
\begin{eqnarray}
y(t)&=& \tan(\alpha_m){L_{\rm x}\over 2}\biggl[1 - \biggl({2x\over L_{\rm x}}\biggr)^2\biggr]v(t) + C \label{eq_y}
\end{eqnarray}
where $x$ is the position of the wall particle with respect to the horizontal axis (assuming that $x=0$ for the center of the
top/bottom wall) and $v(t) = \gamma L_{\rm x}$; the shear rate $\gamma=10^{-4}\pi\cos(2\pi t/T)/(2L_{\rm x})\approx10^{-6}$.
The constant $C$ assumes the appropriate value for the top and bottom wall particles.

We let the top wall slide up and down to readjust the pressure
until the system reaches a stationary shear cycle. Then, we fix the end points
of both walls and continue shearing until the pressure inside of the
system (found from Cauchy stress tensor), averaged over a shear cycle,
reaches a constant value. The animation is available as supplementary material~\cite{supp_movie}.

In the discussion that follows, we will use the average value of the pressure
on top and bottom wall exerted by the system particles, $P_1$
and $P_2$, respectively, to define the confining pressure,
$P_{\rm cf}=(P_1+P_2)/2$.

\subsection{Energies in Sheared Granular Assembly}
\label{app_energies}
Here we discuss the energy balance during shear, considering in detail
energy input, dissipation, and balance.

\subsubsection{Energy Input}
During shear, the top and bottom walls have a prescribed parabolic
shape that changes over time, and left and right boundaries are
fixed. To compute the energy that is added to the system by moving the
walls, one has to integrate the force over the boundary.  There is no
energy added to the system through the (fixed) left and right wall and
we only need to find the energy entering through the collision of
system particles with the top and bottom wall.  This energy is given
by
\begin{eqnarray}
  E_{\rm w}=\sum_{\rm j}\boldsymbol F_{\rm j}\cdot \boldsymbol n_{\rm j} ds\,\label{eq_ew} ,
\end{eqnarray}
where $j$ sums over all collisions of the bottom and top wall
particles with any of the system particles.  Here, $\boldsymbol n_{\rm j}$
is the unit vector normal to the boundary at the location of the wall
particle experiencing a collision, ${\bf F}_{\rm j}$ is the force on
the wall particle and $ds$ is the length element (here the wall
particle diameter).

For the direction normal to the boundary at the center of the particle
$w_{\rm j}$, we have $\boldsymbol n_{\rm j} \cdot \boldsymbol t_{\rm j}=0$ with
$\boldsymbol t_{\rm j}$ being a unit vector tangential to the boundary at
$w_{\rm j}$. The slope of the curve with the tangent vector
$\boldsymbol t_{\rm j}$ is given by $y|_{x=x_{\rm j}}=2ax_{\rm j}$, where the
value of $a$ is obtained from $y(t)$ and $x=x_{\rm j}$ in Eq.~(\ref{eq_y}). Thus
$\boldsymbol t_{\rm j}=(\boldsymbol{e}_{\rm x}+2ax_{\rm j}
\boldsymbol{e}_{\rm y})/(1+4a^2x_{\rm j}^2)^{1/2}$ and the unit normal vector
$\boldsymbol n_{\rm j}$ is given by
\begin{eqnarray}
  \boldsymbol n_{\rm j}=\frac{(-2ax_{\rm j}\boldsymbol{e}_{\rm x} + \boldsymbol{e}_{\rm y})}{\sqrt{1+4a^2x_{\rm j}^2}}\label{eq_n}
\end{eqnarray}
Finally, from Eq.~(\ref{eq_ew}) and Eq.~(\ref{eq_n}) we obtain the expression for the total energy added to the
system by the moving walls
\begin{eqnarray}
  E_{\rm w} &=&\sum_{\rm j}\frac{(-2ax_{\rm j}\boldsymbol{e}_{\rm x}+\boldsymbol{e}_{\rm y})}{\sqrt{1+4a^2x_{\rm j}^2}}\cdot \boldsymbol F_{\rm j} ds\, .
\end{eqnarray}

\subsubsection{Calculations of the Relevant Energy Contributions}

Total energy stored in capillary bridge(s) between the
particles $i,j$, can be found by integrating the force between them
over the separating distance $S$ (smaller than the maximum
separating distance $S^{ \rm max}$)
\begin{equation}
  E_{\rm c}^{ \rm i,j}= \int_0^{\rm S}|\boldsymbol F_{\rm c,ij}|dS
\end{equation}
with the functional form of $|\boldsymbol F_{\rm c,ij}|$ given in
Eq.~(\ref{eq:cohesive_force}) with the closed form of the energy stored in
a capillary bridge given in Eq.~(\ref{eq_ec}).
The total energy stored in all capillary bridges is
\begin{equation}
  E_{\rm c} = \sum_{\rm i,j}E_{\rm c}^{ \rm i,j}\, ,
\end{equation}
for all pairs of particles $i,j$ that interact via capillary force.
\vspace{0.2cm}

Kinetic energy is computed as
\begin{equation}
 E_{\rm k}=\sum_{\rm i=1}^{\rm N} \frac{m_{\rm i}|\boldsymbol v_{\rm i}|^2}{ 2}\, ,
\end{equation}
where $N$ is the total number of system particles and
$m_{\rm i},~|\boldsymbol v_{\rm i}|$ are the mass and velocity of
the $i$--th particle, respectively.

Elastic energy is computed as
\begin{equation}
  E_{\rm el}=\sum_{\rm i,j} \frac{k_{\rm n}x_{\rm i,j}^2}{2}\, ,
\end{equation}
where $i,j$ runs over all pairs of overlapping particles, including
the system particle -- wall particle interactions.

\subsubsection{Energy Balance}

Energy dissipated due to rupturing of capillary bridges is equal to
the capillary energy at the maximum separating distance
$S^{\rm max}$.  To find the total energy dissipated due to
breaking and reformation of the bridges, we sum over all ruptured
bridges
\begin{eqnarray}
  E_{\rm bb}&=&\sum_{\rm i,j} \int_0^{S^{\rm max}}|\boldsymbol F_{\rm c,ij}|dS \,.
\end{eqnarray}
The indices $i,j$ refer to all pairs of particles that experienced bridge rupturing.
\vspace{0.2cm}

Total energy dissipated during a time step can be computed from the
 energy balance equation described next. The energy entering the system due
to the moving walls has to be equal to the sum of the changes in the
elastic, kinetic and capillary energies,
$\Delta E_{\rm el}(t), \Delta E_{\rm k}(t)$ and
$\Delta E_{\rm b}(t) $, between two consecutive time steps,
$(t-\Delta t)$ and $ t$, the energy dissipated by breaking and
reformation of capillary bridges $E_{\rm bb}(t)$ and the energy
dissipated due to friction and other non--linear effects,
$E_{\rm l}(t)$.  The balance equation takes the form
 \begin{equation}
  E_{\rm w}(t) = \Delta E_{\rm el}(t) + \Delta E_{\rm k} (t)  + \Delta E_{\rm c} (t) + E_{\rm l}(t) +  E_{ \rm bb}(t)\, .    \label{eq:energy}
 \end{equation}
From Eq.~(\ref{eq:energy}), we can compute the dissipated energy,
$E_{\rm diss}(t) = E_{\rm l} + E_{\rm bb}$. For the dry systems we can use the same equation to find $E_{\rm diss}$ --
there is no cohesion in the system and $E_{\rm c}, ~E_{\rm bb} = 0$ trivially.
Note that we ignore the energy dissipation due to viscous effects, as
appropriate for the slow shear rates
considered~\cite{Herminghaus2005, Karmakar_unpub}.

\section{Results}
\label{sec:results}


\subsubsection{Pressure Evolution}

 \begin{figure}
  \centering
   \includegraphics[width=2.3in]{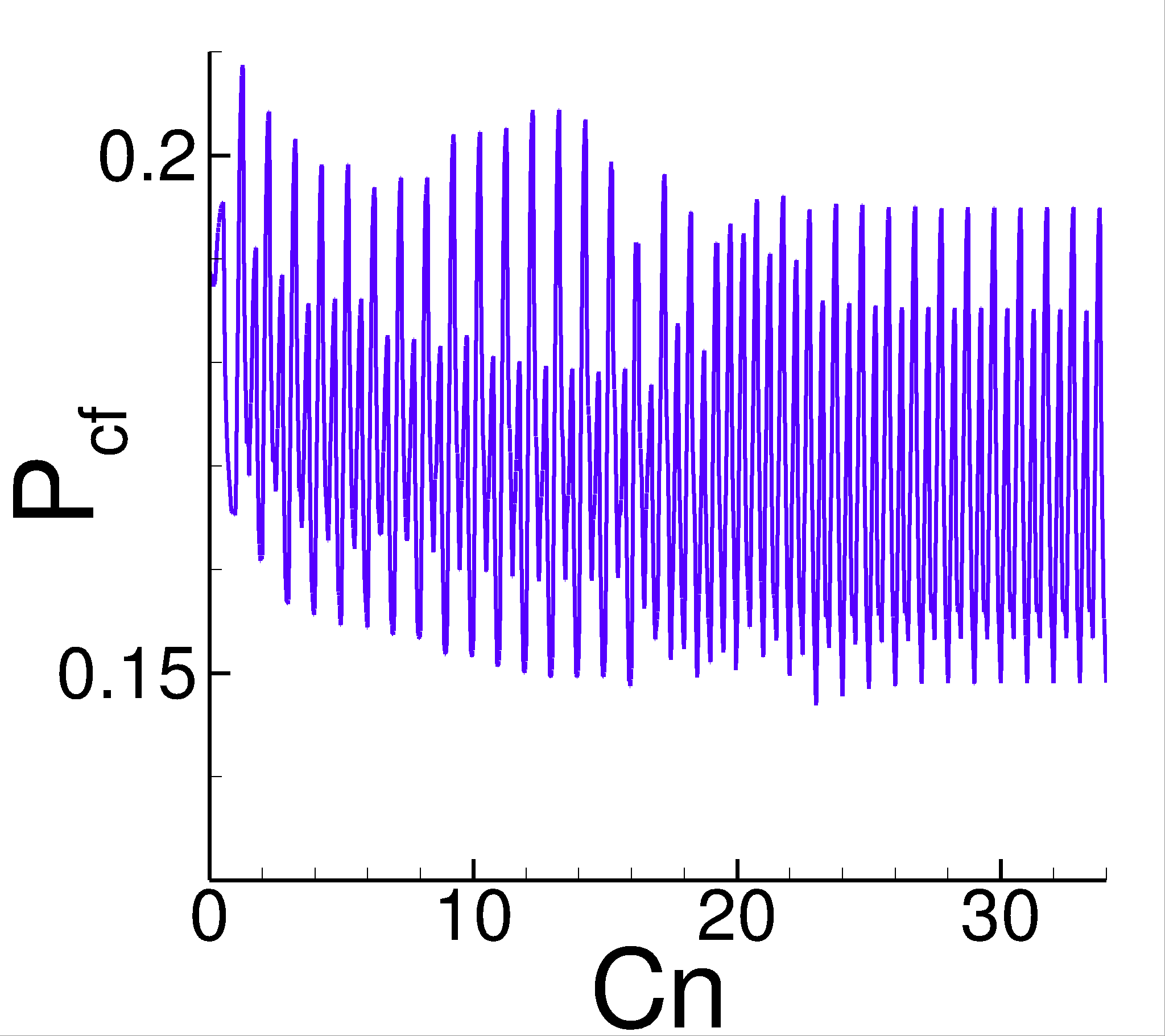}
   \caption{Evolution of the average confining pressure, $P_{\rm cf}$, over the
   cycles (denoted by $C_{\rm n}$) for wet system for the initial
   pressure $P_{\rm init}\approx 0.2$.}
   \label{fig:PabsPdiff}
 \end{figure}

Figure~\ref{fig:PabsPdiff} shows the average pressure on the top and bottom wall, $P_{\rm cf}$, for the wet systems with $P_{\rm init}\approx 0.2$.
As already mentioned in  Sec.~\ref{sec:methods}, $P_{\rm cf}$ evolves during first couple of shear cycles and
finally reaches a stable behavior after $C_{\rm n}\approx 20$. Therefore the data used
to draw our conclusions presented in this section are collected only after $P_{\rm cf}$ stabilizes.
Note that the number of cycles needed to achieve stable $P_{\rm cf}$ behavior differs for each value of $P_{\rm init}$
and we average the results over last $15$ (stable) cycles.

As a side note, we comment that
Fig.~\ref{fig:PabsPdiff} shows that pressure evolution is not symmetric during a cycle, even when a stable regime
is reached: the peak in $P_{\rm cf}$ that occurs when the system is sheared up towards $\alpha_{\rm m}$ is larger than the
peak in $P_{\rm cf}$ when the system is sheared down towards $-\alpha_{\rm m}$. This
asymmetry is observed for all different $P_{\rm init}$ and is due to the fact that we
start shearing towards $\alpha_{\rm m}$ initially. We verified that shearing initially
in the opposite direction reverses the asymmetry.    While the influence of the initial conditions
is not the focus of this paper, an existence of such a long-time memory of the system appears as a topic
that requires further research.

\subsubsection{Energy Transfer}

\begin{figure}[tb!]
    \includegraphics[width=0.8\columnwidth]{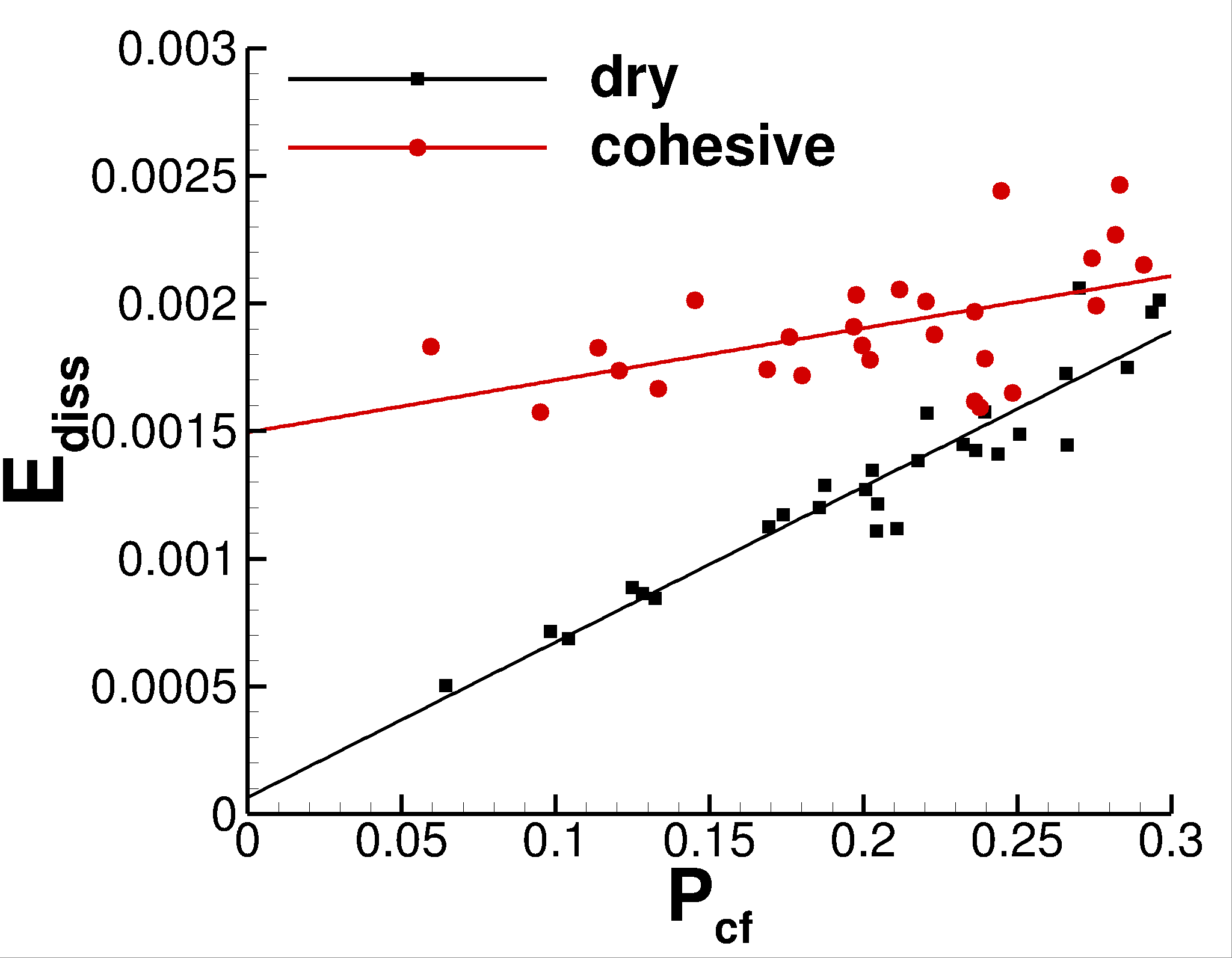}
  \caption{Dissipated energy during shear as a function of the confining
    pressure for dry (black squares) and wet particles (red circles). The results are
    averaged over 15 cycles. For the wet system,
    $E_{\rm diss}=E_{\rm l}+E_{\rm bb}$, see
    Eq.~(\ref{eq:energy}).
    }\label{fig:energy_dissipation}
\end{figure}

Equation~(\ref{eq:energy}) allows to compute the energy dissipation
for both wet and dry particles.
Figure~\ref{fig:energy_dissipation} shows the total
dissipated energy,
$E_{\rm diss}=E_{\rm l}+E_{\rm bb}$ (recall that in the dry case~$E_{\rm bb}$ is
trivially zero), averaged
over $15$ stationary shear cycles, for both wet and dry assemblies.
We note that the simulations, as implemented, are limited
in the range of pressures that can be considered.  For
$P_{\rm cf} \lessapprox 0.06$ it is difficult to carry out
simulations since the particles may detach from the walls.  For
pressures larger than $P_{\rm cf}\gtrapprox0.3$, the overlap between
the particles becomes large, suggesting that a different
interaction model may be needed there.

Figure~\ref{fig:energy_dissipation} shows that the energy dissipated
by friction increases  with the confining pressure, $P_{\rm cf}$, in a manner which is consistent with linear behavior (although the scatter of data is significant, particularly for the wet systems). The increase of dissipated energy with $P_{\rm cf}$ is steeper for dry granulate compared to the wet one, as expected since the coefficient of friction for dry particles is larger.
Further simulations that are beyond the scope of this work will be needed to
decrease the scatter of the data and confirm the ratio of the slopes.

The trend of the results shown in Fig.~\ref{fig:energy_dissipation}
clearly suggests that for the wet particles, the energy dissipated for
$P_{\rm cf}\approx 0$, has a non--zero value, while
$E_{\rm diss} \approx 0$ for the dry case.  For the larger pressures we notice
that the energy is dissipated at a similar level for both types of considered
systems.

\subsubsection{Non-affine Motion}

To investigate the non-zero energy loss at small confining pressure in the wet systems we
focus next on the energy dissipation via breaking and reforming of the capillary
bridges. The amount of the
energy dissipated by breaking bridges per shear cycle can be
computed directly from the number of capillary bridges as a function of time.

\begin{figure}[tb!]
    \includegraphics[width=0.8\columnwidth]{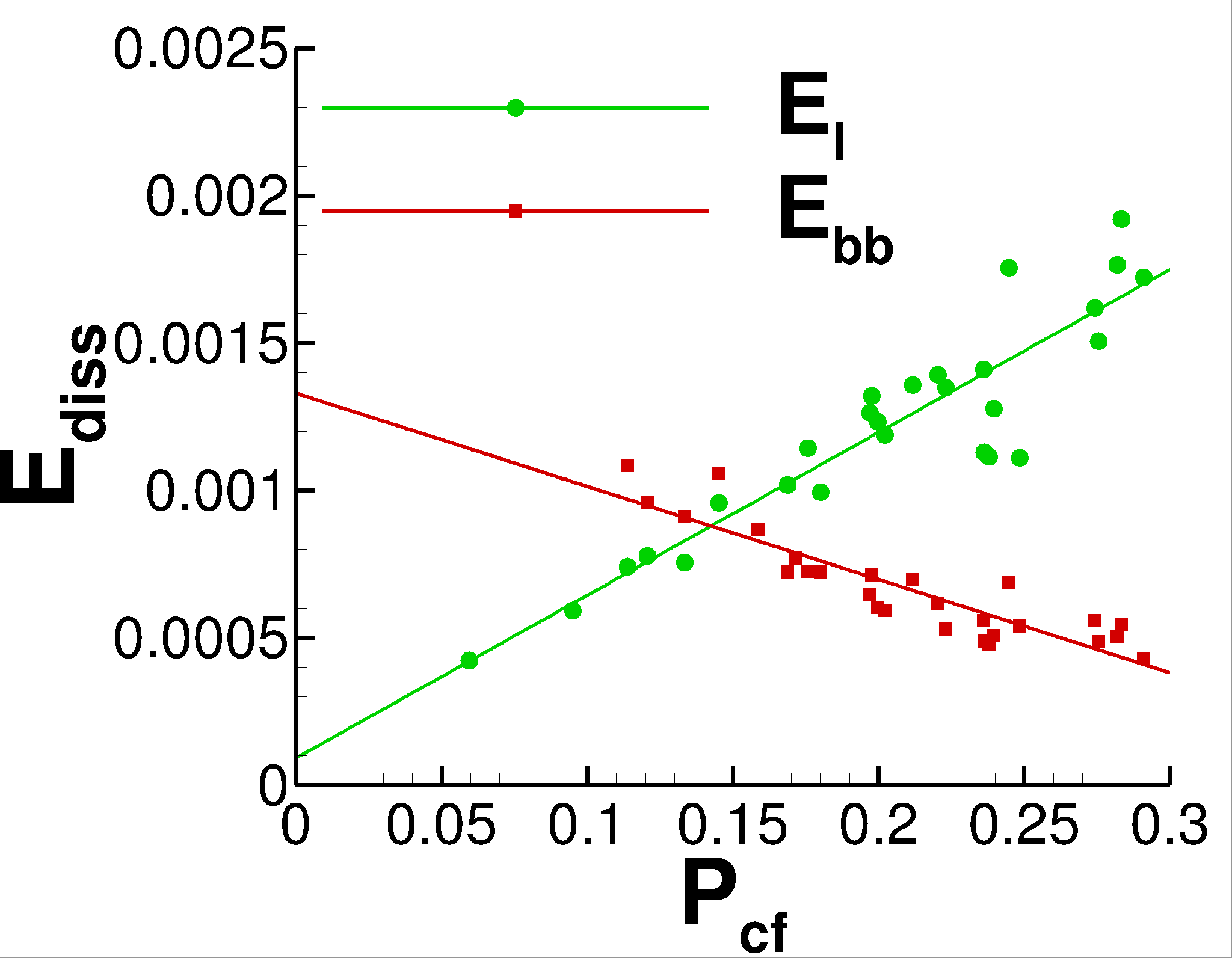}
      \caption{
    Energy dissipation from breaking and reformation of the bridges
    and other non--linear effects as a function of $P_{\rm cf}$ averaged
    over $15$ cycles for each $P_{\rm cf}$.}
    \label{numerics:breakingBridges_a}
    \end{figure}

Figure~\ref{numerics:breakingBridges_a} shows the energy dissipated by
breaking and reforming of the capillary bridges, $E_{\rm bb}$,
and by friction and inelastic collisions, $E_{\rm l}$. We
observe that there is a crossover between the regime where the energy
is dissipated mainly by $E_{\rm bb}$ for small
$P_{\rm cf}$, and mostly from friction and inelastic collisions, in $E_{\rm l}$,
for sufficiently large $P_{\rm cf}$. We note that the energy
dissipated in $E_{\rm bb}$ is decreasing
with the increasing value of $P_{\rm cf}$. This finding can be
rationalized as follows: for  lager $P_{\rm cf}$, particles have
less space to move and therefore there are fewer bridges that
break. To support this explanation, we consider affinity of particle motion.

During shear, the particles move not only in the manner imposed by the
moving walls, but also relative to each other. The relative motion of
the particles, also referred to as the non--affine motion, is expected to
play a role with regards to the breaking and reformation of the
capillary bridges and subsequently the energy dissipation tied to the
capillary effects, $E_{\rm bb}$.
Therefore, with the particular goal of explaining the decrease of
$E_{\rm bb}$ with increasing $P_{\rm cf}$ (see Figure~\ref{numerics:breakingBridges_a})
we investigate the non--affine motion of the
particles as a function of $P_{\rm cf}$,
using the approach described in~\cite{pre12_impact}, and outlined
briefly here.

First, for every particle $p$, we find the affine deformation matrix $A_{\rm p}(t)$ at
the time $t$ with the property
\begin{eqnarray}
  A_{\rm p}(t)\boldsymbol{\rm{r}}_{\rm p}(t) = \boldsymbol{\rm r}_{\rm p}(t+\delta t)\, ,
\end{eqnarray}
where $\boldsymbol{\rm r}_{\rm p}(t)$ is the position of the particle
$p$. The non--affine motion is defined as the
minimum of the mean squared displacement
\begin{eqnarray}
  D^{2}_{\rm min} = {\rm min}\Biggl\{\sum_{n=1}^m||\boldsymbol{\rm r}_{\rm n} - \boldsymbol{\rm r}_{\rm p} - [A_{\rm n}\boldsymbol{\rm r}_{\rm n} - A_{\rm p}\boldsymbol{\rm r}_{\rm p}]||^2\Biggr\}\, ,
\end{eqnarray}
where $m$ is the number of particles within the distance of
$2.5d_{\rm ave}$ from the particle $p$, and
$\boldsymbol{\rm r}_{\rm n}(t)$ is the position of the $n$--th
particle within this distance.

\begin{figure}
  \subfloat[$P_{\rm cf} = 0.11$]{\includegraphics[width = 0.35\textwidth]{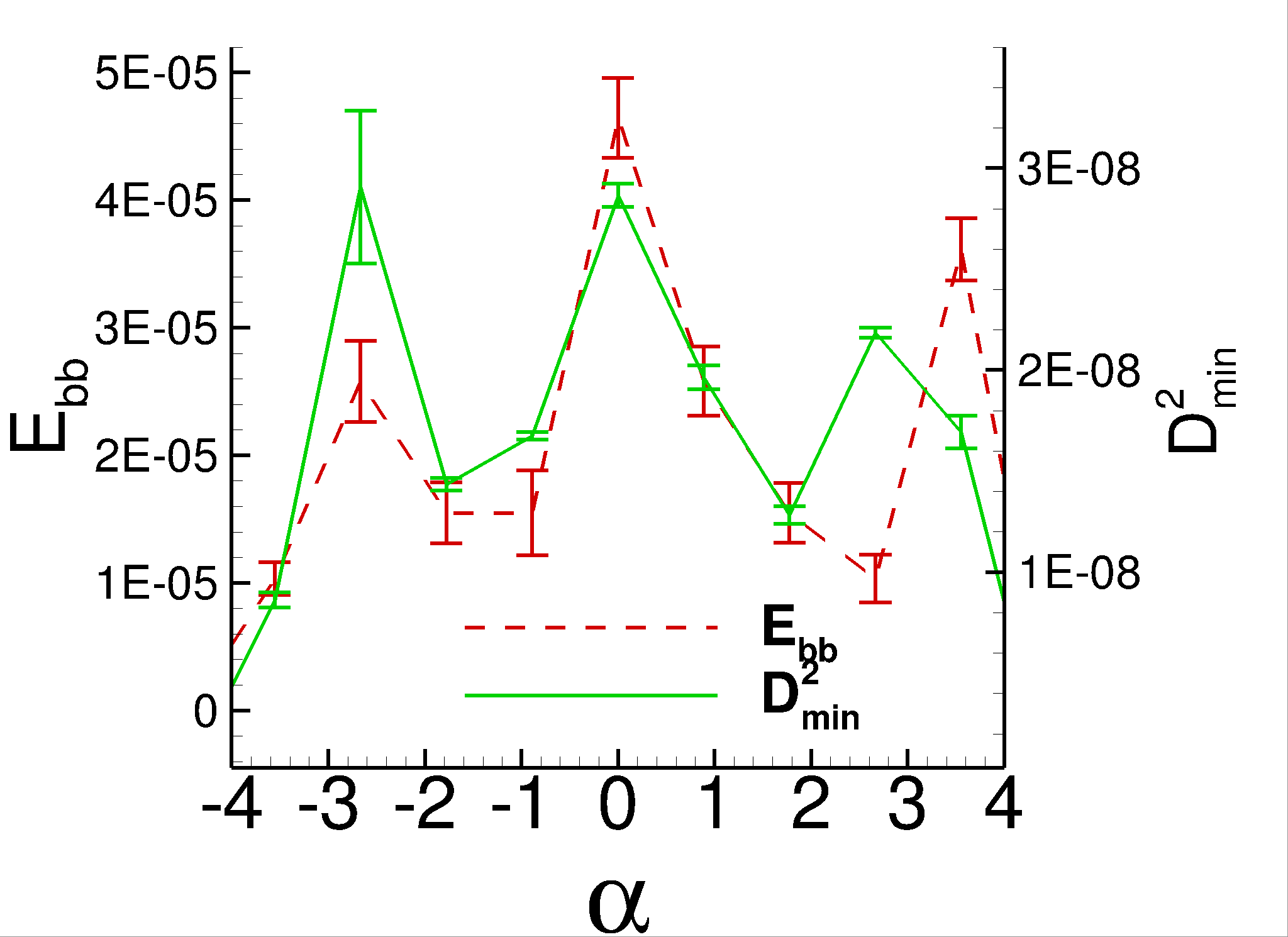}}\\
  \subfloat[$P_{\rm cf} = 0.28$]{\includegraphics[width = 0.35\textwidth]{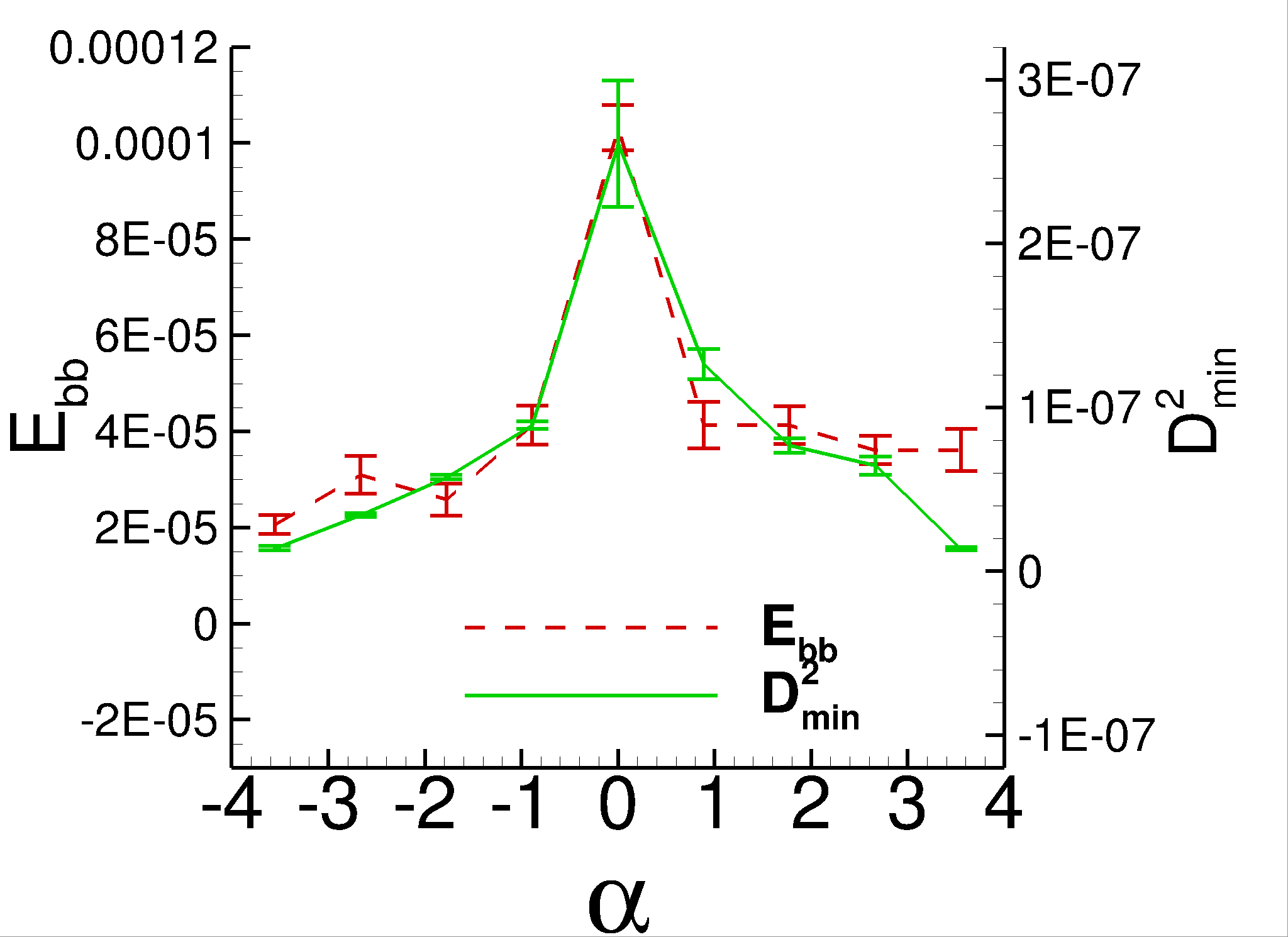}}
  \caption{$E_{\rm bb}$ and $D^2_{\rm min}$ as a function of the shearing angle, $\alpha$, for two different pressures.
  The results are averaged over $15$ cycles.}\label{fig:bridges_nonaffine}
\end{figure}

Figure~\ref{fig:bridges_nonaffine} shows the energy loss from the broken bridges, $E_{\rm bb}$,
as well as the measure of non-affine motion, $D^{2}_{\rm min}$, averaged over $15$ cycles,
as a function of the shearing angle, $\alpha$.
We average the results over $9$ equal size segments between $-\alpha_{\rm m}$ and $\alpha_{\rm m}$; we choose
this number of segments so to able to show trends while still having reasonable statistics.
The results are shown for one small and for
one large confining pressure, $P_{\rm init}=0.08$ and $P_{\rm init}=0.4$, respectively.
We see a clear correlation between $E_{\rm bb}$
and $D^{2}_{\rm min}$; similar correlation is seen for other values of $P_{\rm init}$.
For $P_{\rm init }=0.08$, the magnitude of $D^2_{\rm min}$ is much higher than for $P_{\rm init}=0.4$;
the particles have more freedom to move in non-affine manner for small confining pressures.
Furthermore, the value of $D^{2}_{\rm min}$ is largest for $\alpha \approx 0$, when the shearing
speed assumes its maximum.

\begin{figure}[tb!]
    \includegraphics[width=0.8\columnwidth]{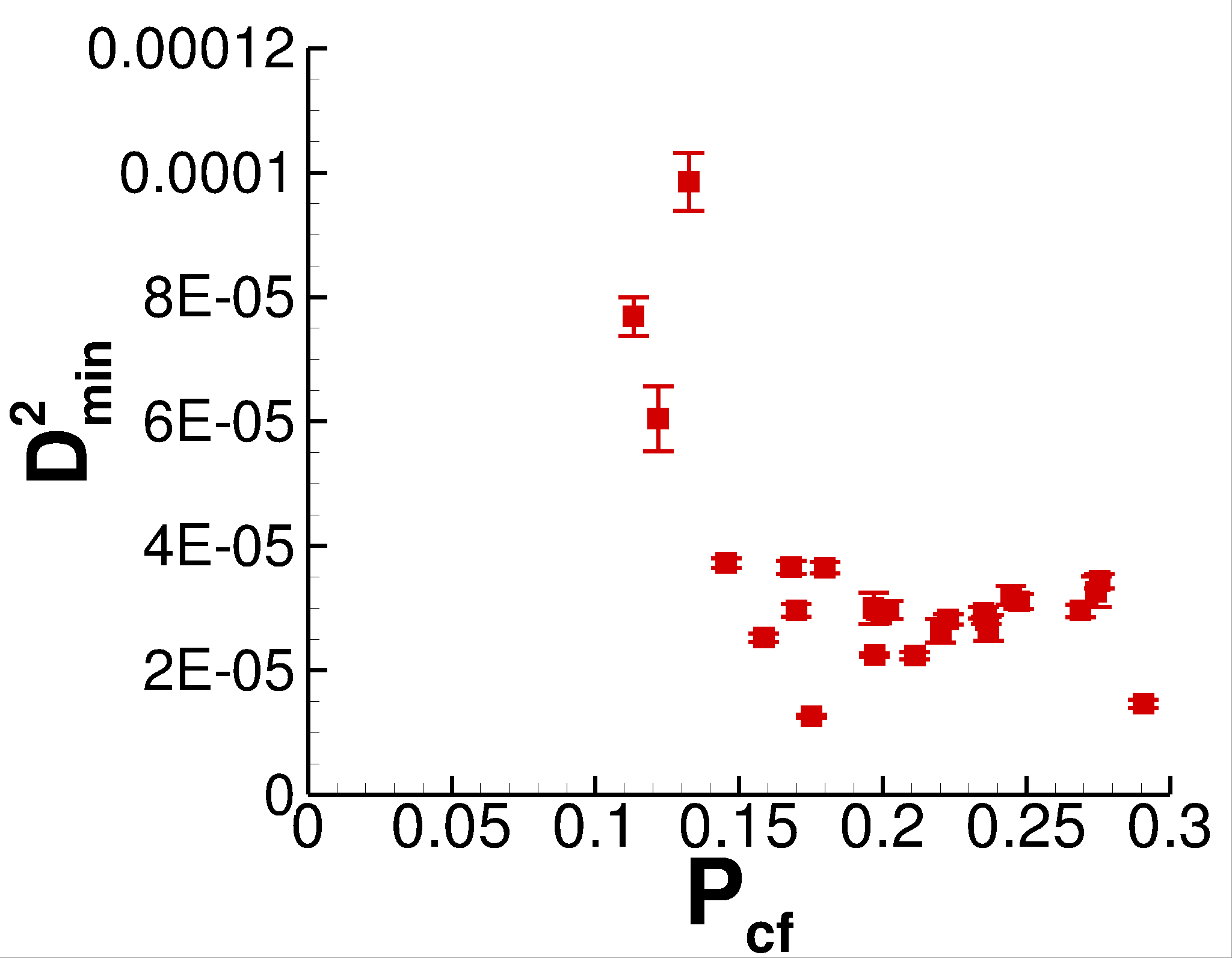}
  \caption{Non--affine motion in the wet
    systems averaged over $15$ cycles
    for each $P_{\rm cf}$.}\label{numerics:breakingBridges_b}
\end{figure}

Figure~\ref{numerics:breakingBridges_b} summarizes the results related non-affine motion.  Here we plot
non--affine motion as a function of $P_{\rm cf}$, averaged over
time and over~$15$ cycles. An overall decreasing trend of the total
non-affine motion is obvious.  As a side note, we comment that we have verified that a modest increase
in particle stiffness does not influence the trend of the non--affine
motion with changing pressure.

The above findings show that the degree of non-affine motion is directly connected
to the breaking and reforming of the bridges and reversely to the confining pressure, $P_{\rm cf}$.
Therefore, a decrease of the non-affine motion explains the decrease of $E_{\rm bb}$ with increasing $P_{\rm cf}$.

We note that we have also computed non--affine motion for
the  frictionless and elastic particles.  The results (figure not included for brevity)
show significantly larger degree of non-affinity for both considered systems,
compared to the frictional, inelastic one considered so far. This is expected
due to reduced energy loss in elastic and/or frictionless systems.

\subsubsection{Internal Cohesion}

Another manner in which cohesion can influence energy dissipation is
internal cohesion: capillary bridges pull particles together, leading
to an enhanced friction at the contacts~\cite{Fournier_05,Mason_99} as well as damping due to
inelasticity (see Appendix~\ref{app_sim}) that may influence the
energy dissipation through changing $E_{\rm l}$ in
Eq.~(\ref{eq:energy}). To compute this effect, we proceed as
follows: In equilibrium, the capillary force between the particles
leads to a compressive force and an overall elastic energy
$E_{\rm el}\approx 10^{-6}$. Then, we carry out simulations
with dry systems and find that this value of elastic energy
corresponds to $P_{\rm cf}\approx 0.07$. This value of
$P_{\rm cf}$ leads to
$E_{\rm l} \approx 5 \times 10^{-4}$, see
Figure~\ref{numerics:breakingBridges_a}, being less than a third of the
energy dissipated by breaking and reformation of capillary bridges,
$E_{\rm bb}$.  Therefore, the results of our simulations
suggest that the breakup of capillary bridges, and not the friction or
the inelasticity of collisions, is the main source of energy
dissipation in weakly compressed systems and causes non-zero values of $E_{\rm diss}$
as $P_{\rm cf}\rightarrow 0$.

We note that linear extrapolation of the data shown in
Figure~\ref{numerics:breakingBridges_a} to $P_{\rm cf}< 0.06$
suggests even smaller values of $E_{\rm l}$; however since
simulations can not be reliably carried out for such smaller values,
we conservatively choose the value of
$5 \times 10^{-4}$ as the upper bound for $E_{\rm l}$ for
$P_{\rm cf} \approx 0$. The main point, that only a small
part of energy is lost due to friction and inelasticity for small
confining pressures, is clear from the overall trend of the data.

\section{Conclusion}
\label{sec:conclusion}

In this paper, we discussed the origin of energy dissipation
in sheared wet and dry granular systems in numerical simulations
following a similar setup as in recent
experiments~\cite{Herminghaus2005, Fournier_05} in the regime characterized
by the presence of individual capillary bridges. For small
confining pressure, wet systems are stiffer than dry ones due
to the cohesion by virtue of capillary bridges formed between
neighboring particles, which increase the energy dissipation in two
ways.  For the material parameters used in the simulations,
about two thirds of the dissipated energy was found to result
from breaking of the capillary bridges which were elongated above their
maximum length. The remaining one third is caused by the cohesion
which increases the contact forces between particles and thus causes
friction even in an unconfined wet granulate.  An increase of applied
confining pressure is found to have two consequences.  First, the
simulations show that energy dissipation due to breakup of capillary
bridges becomes less relevant due to a decrease of the non-affine
particle motion.  Second, the energy dissipation for both dry and wet
granulates increases linearly with externally applied confining
  pressure due to increased particle--particle friction. Thus for
sufficiently large confining pressure energy dissipation is always
dominated by friction between grains. We expect that the exact
proportions describing relevance of various energy loss mechanisms
may depend on the material parameters of the granular particles,
however the main conclusion is general: for sufficiently small pressures,
the dominant part of the energy loss is due to the effects related
to cohesion, and in particular due to breakups of capillary bridges.

\begin{acknowledgments}
  Stimulating discussions with Stephan Herminghaus are gratefully
  acknowledged.  We further acknowledge the support by: the German
  Research Foundation (DFG) via GRK - 1276, Saarland University (SK,
  RS) and SPP 1486 ``PiKo'' under Grant No. HE 2016/14-2 (MB); by a
  start up grant from Saarland University (ALS); and by the NSF Grant
  No. DMS- 1521717 and DARPA contract No. HR0011-16-2- 0033 (LK, LK).
\end{acknowledgments}

\appendix
\section{Details of Simulation Techniques}
\label{app_sim}

The particles in the considered numerical system are modeled as 2D
soft frictional inelastic circular particles that interact via normal and
tangential forces, specified here in nondimensional form, using
${\bar d}, ~m,~\tau_{\rm c}$ as the length, mass and time scale introduced in
the main text of the paper.

Dimensionless normal force between $i$--th and $j$--th particle is
\begin{equation}
  \boldsymbol F_{\rm i,j}^{\rm n}=k_{\rm n} x_{\rm i,j}\boldsymbol n
  -\eta_{\rm n}\overline m\boldsymbol v_{\rm i,j}^{\rm n}\, ,
\end{equation}
where $\boldsymbol v_{\rm i,j}^{\rm n}$ is the relative normal velocity,
$\overline{m}$ is reduced mass,
$x_{\rm i,j} = d_{\rm ave} - r_{\rm i,j}$ is the amount of
compression, with $d_{\rm ave} = {(d_{\rm i} + d_{\rm j})/2}$ and
$d_{\rm i}$, $d_{\rm j}$ diameters of the particles $i$ and $j$. The
distance of the centers of $i$--th and $j$--th particle is denoted as
$r_{\rm i,j}$. Parameter $\eta_{\rm n}$ is the damping coefficient
in the normal direction, related to the coefficient of restitution
$e$.

We implement the Cundall--Strack model for static friction
\cite{cundall79}. The tangential spring $\boldsymbol\xi$ is
introduced between particles for each new contact that forms at time
$T = T_0$ and is used to determine the tangential force during the
contact of particles. Due to the relative motion of particles, the
spring length $\xi$ evolves as
$\xi=\int_{T_0}^T\boldsymbol
v_{\rm i,j}^{\rm t}(t)dt$ with
$\boldsymbol v_{\rm i,j}^{\rm t}=\boldsymbol v_{\rm i,j}-\boldsymbol
v_{\rm i,j}^{\rm n}$ and $\boldsymbol v_{\rm i,j}$ being the relative
velocity of particles $i,~j$. The tangential direction is defined as
$\boldsymbol{t}={\boldsymbol v_{\rm i,j}^{\rm t}/|\boldsymbol
  v_{\rm i,j}^{\rm t}|}$. The direction of $\boldsymbol\xi$
evolves over time and we thus correct the tangential spring as
$\boldsymbol{\xi}^{\prime}=\boldsymbol{\xi}
-\boldsymbol{n}(\boldsymbol{n.\xi})$. The tangential force is
set to
\begin{equation}
  \boldsymbol F^{\rm t} =
  \min(\mu_{\rm s} |\boldsymbol F^{\rm n} |, |\boldsymbol F^{\rm t\ast} |)
  {\boldsymbol F^{\rm t\ast} /|\boldsymbol F^{\rm t\ast} | }\, ,
\end{equation}
with
\begin{equation}
  \boldsymbol F^{\rm t\ast} =
  -k_{\rm t} \boldsymbol \xi^{\prime}
  - \eta_{\rm t}  m\boldsymbol v_{\rm i,j}^{\rm t}\, .
\end{equation}
Viscous damping in the tangential direction is included in the model
via the damping coefficient $\eta_{\rm t} = \eta_{\rm n}$.


\begin{thebibliography}{27}
\expandafter\ifx\csname natexlab\endcsname\relax\def\natexlab#1{#1}\fi
\expandafter\ifx\csname bibnamefont\endcsname\relax
  \def\bibnamefont#1{#1}\fi
\expandafter\ifx\csname bibfnamefont\endcsname\relax
  \def\bibfnamefont#1{#1}\fi
\expandafter\ifx\csname citenamefont\endcsname\relax
  \def\citenamefont#1{#1}\fi
\expandafter\ifx\csname url\endcsname\relax
  \def\url#1{\texttt{#1}}\fi
\expandafter\ifx\csname urlprefix\endcsname\relax\def\urlprefix{URL }\fi
\providecommand{\bibinfo}[2]{#2}
\providecommand{\eprint}[2][]{\url{#2}}

\bibitem[{\citenamefont{Faure et~al.}(2001)\citenamefont{Faure, York, and
  Rowe}}]{Faure2001}
\bibinfo{author}{\bibfnamefont{A.}~\bibnamefont{Faure}},
  \bibinfo{author}{\bibfnamefont{P.}~\bibnamefont{York}}, \bibnamefont{and}
  \bibinfo{author}{\bibfnamefont{R.}~\bibnamefont{Rowe}},
  \bibinfo{journal}{Eur. J. Pharm. Biopharm.} \textbf{\bibinfo{volume}{52}},
  \bibinfo{pages}{269 } (\bibinfo{year}{2001}).

\bibitem[{\citenamefont{Leuenberger}(2001)}]{Leuenberger2001A}
\bibinfo{author}{\bibfnamefont{H.}~\bibnamefont{Leuenberger}},
  \bibinfo{journal}{Eur. J. Pharm. Biopharm.} \textbf{\bibinfo{volume}{52}},
  \bibinfo{pages}{279 } (\bibinfo{year}{2001}).

\bibitem[{\citenamefont{Paul and Sharma}(1999)}]{Paul1999}
\bibinfo{author}{\bibfnamefont{W.}~\bibnamefont{Paul}} \bibnamefont{and}
  \bibinfo{author}{\bibfnamefont{C.~P.} \bibnamefont{Sharma}},
  \bibinfo{journal}{J. Mater. Sci. Mater. Med.} \textbf{\bibinfo{volume}{10}},
  \bibinfo{pages}{383 } (\bibinfo{year}{1999}).

\bibitem[{\citenamefont{Komlev et~al.}(2002)\citenamefont{Komlev, Barinov, and
  Koplik}}]{Komlev2002}
\bibinfo{author}{\bibfnamefont{V.}~\bibnamefont{Komlev}},
  \bibinfo{author}{\bibfnamefont{S.}~\bibnamefont{Barinov}}, \bibnamefont{and}
  \bibinfo{author}{\bibfnamefont{E.}~\bibnamefont{Koplik}},
  \bibinfo{journal}{Biomaterials} \textbf{\bibinfo{volume}{23}},
  \bibinfo{pages}{3449 } (\bibinfo{year}{2002}).

\bibitem[{\citenamefont{Simons and Fairbrother}(2000)}]{Simons2000}
\bibinfo{author}{\bibfnamefont{S.}~\bibnamefont{Simons}} \bibnamefont{and}
  \bibinfo{author}{\bibfnamefont{R.}~\bibnamefont{Fairbrother}},
  \bibinfo{journal}{Powder Tech.} \textbf{\bibinfo{volume}{110}},
  \bibinfo{pages}{44 } (\bibinfo{year}{2000}).

\bibitem[{\citenamefont{Litster and Ennis}(2004)}]{Litster2004}
\bibinfo{author}{\bibfnamefont{J.}~\bibnamefont{Litster}} \bibnamefont{and}
  \bibinfo{author}{\bibfnamefont{B.}~\bibnamefont{Ennis}},
  \emph{\bibinfo{title}{Engineering of Granulation Processes: Particle
  Technologies Series, Springer Netherlands}} (\bibinfo{publisher}{Springer
  Netherlands}, \bibinfo{year}{2004}).

\bibitem[{\citenamefont{Realpe and Vel\'{a}zquez}(2008)}]{Realpe2007}
\bibinfo{author}{\bibfnamefont{A.}~\bibnamefont{Realpe}} \bibnamefont{and}
  \bibinfo{author}{\bibfnamefont{C.}~\bibnamefont{Vel\'{a}zquez}},
  \bibinfo{journal}{Chem. Eng. Sci.} \textbf{\bibinfo{volume}{63}},
  \bibinfo{pages}{1602 } (\bibinfo{year}{2008}).

\bibitem[{\citenamefont{Shinagawa and Hirashima}(1998)}]{Shinagawa1998}
\bibinfo{author}{\bibfnamefont{K.}~\bibnamefont{Shinagawa}} \bibnamefont{and}
  \bibinfo{author}{\bibfnamefont{Y.}~\bibnamefont{Hirashima}},
  \bibinfo{journal}{Met. and Mat.} \textbf{\bibinfo{volume}{4}},
  \bibinfo{pages}{350} (\bibinfo{year}{1998}).

\bibitem[{\citenamefont{B.~Bhandari and Schuck}(2013)}]{Bhandari2013}
\bibinfo{author}{\bibfnamefont{M.~Z.} \bibnamefont{B.~Bhandari},
  \bibfnamefont{N.~Bansal}} \bibnamefont{and}
  \bibinfo{author}{\bibfnamefont{P.}~\bibnamefont{Schuck}},
  \emph{\bibinfo{title}{Handbook of Food Powders}}
  (\bibinfo{publisher}{Woodhead Publishing}, \bibinfo{year}{2013}).

\bibitem[{\citenamefont{G\'eminard et~al.}(1999)\citenamefont{G\'eminard,
  Losert, and Gollub}}]{Geminard1999}
\bibinfo{author}{\bibfnamefont{J.-C.} \bibnamefont{G\'eminard}},
  \bibinfo{author}{\bibfnamefont{W.}~\bibnamefont{Losert}}, \bibnamefont{and}
  \bibinfo{author}{\bibfnamefont{J.}~\bibnamefont{Gollub}},
  \bibinfo{journal}{Phys. Rev. E} \textbf{\bibinfo{volume}{59}},
  \bibinfo{pages}{5881} (\bibinfo{year}{1999}).

\bibitem[{\citenamefont{Fournier et~al.}(2005)\citenamefont{Fournier,
  Geromichalos, Herminghaus, Kohonen, Mugele, Scheel, Schulz, Schulz, Schier,
  Seemann et~al.}}]{Fournier_05}
\bibinfo{author}{\bibfnamefont{Z.}~\bibnamefont{Fournier}},
  \bibinfo{author}{\bibfnamefont{D.}~\bibnamefont{Geromichalos}},
  \bibinfo{author}{\bibfnamefont{S.}~\bibnamefont{Herminghaus}},
  \bibinfo{author}{\bibfnamefont{M.~M.} \bibnamefont{Kohonen}},
  \bibinfo{author}{\bibfnamefont{F.}~\bibnamefont{Mugele}},
  \bibinfo{author}{\bibfnamefont{M.}~\bibnamefont{Scheel}},
  \bibinfo{author}{\bibfnamefont{M.}~\bibnamefont{Schulz}},
  \bibinfo{author}{\bibfnamefont{B.}~\bibnamefont{Schulz}},
  \bibinfo{author}{\bibfnamefont{C.}~\bibnamefont{Schier}},
  \bibinfo{author}{\bibfnamefont{R.}~\bibnamefont{Seemann}},
  \bibnamefont{et~al.}, \bibinfo{journal}{J. Phys. Condens. Mat.}
  \textbf{\bibinfo{volume}{17}}, \bibinfo{pages}{477} (\bibinfo{year}{2005}).

\bibitem[{\citenamefont{Mitarai and Nori}(2006)}]{Mitarai2006}
\bibinfo{author}{\bibfnamefont{N.}~\bibnamefont{Mitarai}} \bibnamefont{and}
  \bibinfo{author}{\bibfnamefont{F.}~\bibnamefont{Nori}},
  \bibinfo{journal}{Adv. in Phys.} \textbf{\bibinfo{volume}{55}},
  \bibinfo{pages}{1} (\bibinfo{year}{2006}).

\bibitem[{\citenamefont{Scheel et~al.}(2008)\citenamefont{Scheel, Seemann,
  M.~Brinkmann, Sheppard, Breidenbach, and Herminghaus}}]{Scheel2008A}
\bibinfo{author}{\bibfnamefont{M.}~\bibnamefont{Scheel}},
  \bibinfo{author}{\bibfnamefont{R.}~\bibnamefont{Seemann}},
  \bibinfo{author}{\bibfnamefont{M.~D.~M.} \bibnamefont{M.~Brinkmann}},
  \bibinfo{author}{\bibfnamefont{A.}~\bibnamefont{Sheppard}},
  \bibinfo{author}{\bibfnamefont{B.}~\bibnamefont{Breidenbach}},
  \bibnamefont{and}
  \bibinfo{author}{\bibfnamefont{S.}~\bibnamefont{Herminghaus}},
  \bibinfo{journal}{Nat. Mater.} \textbf{\bibinfo{volume}{7}},
  \bibinfo{pages}{189–193} (\bibinfo{year}{2008}).

\bibitem[{\citenamefont{Fiscina et~al.}(2012)\citenamefont{Fiscina, Pakpour,
  Fall, Vandewalle, Wagner, and Bonn}}]{Fiscina2012}
\bibinfo{author}{\bibfnamefont{J.}~\bibnamefont{Fiscina}},
  \bibinfo{author}{\bibfnamefont{M.}~\bibnamefont{Pakpour}},
  \bibinfo{author}{\bibfnamefont{A.}~\bibnamefont{Fall}},
  \bibinfo{author}{\bibfnamefont{N.}~\bibnamefont{Vandewalle}},
  \bibinfo{author}{\bibfnamefont{C.}~\bibnamefont{Wagner}}, \bibnamefont{and}
  \bibinfo{author}{\bibfnamefont{D.}~\bibnamefont{Bonn}},
  \bibinfo{journal}{Phys. Rev. E} \textbf{\bibinfo{volume}{86}},
  \bibinfo{pages}{020103} (\bibinfo{year}{2012}).

\bibitem[{\citenamefont{Fall et~al.}(2014)\citenamefont{Fall, Weber, Pakpour,
  Lenoir, Shahidzadeh, Fiscina, Wagner, and Bonn}}]{Fall2014}
\bibinfo{author}{\bibfnamefont{A.}~\bibnamefont{Fall}},
  \bibinfo{author}{\bibfnamefont{B.}~\bibnamefont{Weber}},
  \bibinfo{author}{\bibfnamefont{M.}~\bibnamefont{Pakpour}},
  \bibinfo{author}{\bibfnamefont{N.}~\bibnamefont{Lenoir}},
  \bibinfo{author}{\bibfnamefont{N.}~\bibnamefont{Shahidzadeh}},
  \bibinfo{author}{\bibfnamefont{J.}~\bibnamefont{Fiscina}},
  \bibinfo{author}{\bibfnamefont{C.}~\bibnamefont{Wagner}}, \bibnamefont{and}
  \bibinfo{author}{\bibfnamefont{D.}~\bibnamefont{Bonn}},
  \bibinfo{journal}{Phys. Rev. Lett.} \textbf{\bibinfo{volume}{112}},
  \bibinfo{pages}{175502} (\bibinfo{year}{2014}).

\bibitem[{\citenamefont{Brown et~al.}(2010)\citenamefont{Brown, Rodenberg,
  Amend, Mozeika, Steltz, Zakin, Lipson, and Jaeger}}]{Brown2010}
\bibinfo{author}{\bibfnamefont{E.}~\bibnamefont{Brown}},
  \bibinfo{author}{\bibfnamefont{N.}~\bibnamefont{Rodenberg}},
  \bibinfo{author}{\bibfnamefont{J.}~\bibnamefont{Amend}},
  \bibinfo{author}{\bibfnamefont{A.}~\bibnamefont{Mozeika}},
  \bibinfo{author}{\bibfnamefont{E.}~\bibnamefont{Steltz}},
  \bibinfo{author}{\bibfnamefont{M.}~\bibnamefont{Zakin}},
  \bibinfo{author}{\bibfnamefont{H.}~\bibnamefont{Lipson}}, \bibnamefont{and}
  \bibinfo{author}{\bibfnamefont{H.}~\bibnamefont{Jaeger}},
  \bibinfo{journal}{Proc. Natl. Acad. Sci.} \textbf{\bibinfo{volume}{107}},
  \bibinfo{pages}{18809} (\bibinfo{year}{2010}).

\bibitem[{\citenamefont{Halsey and Levine}(1998)}]{Halsey1998}
\bibinfo{author}{\bibfnamefont{T.}~\bibnamefont{Halsey}} \bibnamefont{and}
  \bibinfo{author}{\bibfnamefont{A.}~\bibnamefont{Levine}},
  \bibinfo{journal}{Phys. Rev. Lett.} \textbf{\bibinfo{volume}{80}},
  \bibinfo{pages}{3141} (\bibinfo{year}{1998}).

\bibitem[{\citenamefont{Herminghaus}(2005)}]{Herminghaus2005}
\bibinfo{author}{\bibfnamefont{S.}~\bibnamefont{Herminghaus}},
  \bibinfo{journal}{Adv. Phys.} \textbf{\bibinfo{volume}{54}},
  \bibinfo{pages}{221} (\bibinfo{year}{2005}).

\bibitem[{\citenamefont{Karmakar et~al.}()\citenamefont{Karmakar, Schaber,
  Schuhmacher, Scheel, DiMichiel, Brinkmann, Seemann, Kovalcinova, and
  Kondic}}]{Karmakar_unpub}
\bibinfo{author}{\bibfnamefont{S.}~\bibnamefont{Karmakar}},
  \bibinfo{author}{\bibfnamefont{M.}~\bibnamefont{Schaber}},
  \bibinfo{author}{\bibfnamefont{A.-L.} \bibnamefont{Schuhmacher}},
  \bibinfo{author}{\bibfnamefont{M.}~\bibnamefont{Scheel}},
  \bibinfo{author}{\bibfnamefont{M.}~\bibnamefont{DiMichiel}},
  \bibinfo{author}{\bibfnamefont{M.}~\bibnamefont{Brinkmann}},
  \bibinfo{author}{\bibfnamefont{R.}~\bibnamefont{Seemann}},
  \bibinfo{author}{\bibfnamefont{L.}~\bibnamefont{Kovalcinova}},
  \bibnamefont{and} \bibinfo{author}{\bibfnamefont{L.}~\bibnamefont{Kondic}},
  \bibinfo{note}{unpublished}.

\bibitem[{\citenamefont{Kovalcinova et~al.}(2015)\citenamefont{Kovalcinova,
  Goullet, and Kondic}}]{kovalcinova_percolation}
\bibinfo{author}{\bibfnamefont{L.}~\bibnamefont{Kovalcinova}},
  \bibinfo{author}{\bibfnamefont{A.}~\bibnamefont{Goullet}}, \bibnamefont{and}
  \bibinfo{author}{\bibfnamefont{L.}~\bibnamefont{Kondic}},
  \bibinfo{journal}{Phys. Rev. E} \textbf{\bibinfo{volume}{92}},
  \bibinfo{pages}{032204} (\bibinfo{year}{2015}).

\bibitem[{\citenamefont{Kondic}(1999)}]{kondic_99}
\bibinfo{author}{\bibfnamefont{L.}~\bibnamefont{Kondic}},
  \bibinfo{journal}{Phys. Rev. E} \textbf{\bibinfo{volume}{60}},
  \bibinfo{pages}{751} (\bibinfo{year}{1999}).

\bibitem[{\citenamefont{Willett et~al.}(2000)\citenamefont{Willett, Adams,
  Johnson, and Seville}}]{Willett2000}
\bibinfo{author}{\bibfnamefont{C.}~\bibnamefont{Willett}},
  \bibinfo{author}{\bibfnamefont{M.}~\bibnamefont{Adams}},
  \bibinfo{author}{\bibfnamefont{S.}~\bibnamefont{Johnson}}, \bibnamefont{and}
  \bibinfo{author}{\bibfnamefont{J.}~\bibnamefont{Seville}},
  \bibinfo{journal}{Langmuir} \textbf{\bibinfo{volume}{16}},
  \bibinfo{pages}{9396} (\bibinfo{year}{2000}).

\bibitem[{\citenamefont{Urso et~al.}(1999)\citenamefont{Urso, Lawrence, and
  Adams}}]{Urso1999}
\bibinfo{author}{\bibfnamefont{M.~E.~D.} \bibnamefont{Urso}},
  \bibinfo{author}{\bibfnamefont{C.~J.} \bibnamefont{Lawrence}},
  \bibnamefont{and} \bibinfo{author}{\bibfnamefont{M.~J.} \bibnamefont{Adams}},
  \bibinfo{journal}{J. Colloid Interface Sci.} \textbf{\bibinfo{volume}{220}},
  \bibinfo{pages}{42} (\bibinfo{year}{1999}).

\bibitem[{sup()}]{supp_movie}
\emph{\bibinfo{title}{Animation of the shear cycle}}, \urlprefix\url{URL
  provided/inserted by editor}.

\bibitem[{\citenamefont{Kondic et~al.}(2012)\citenamefont{Kondic, Fang, Losert,
  O'Hern, and Behringer}}]{pre12_impact}
\bibinfo{author}{\bibfnamefont{L.}~\bibnamefont{Kondic}},
  \bibinfo{author}{\bibfnamefont{X.}~\bibnamefont{Fang}},
  \bibinfo{author}{\bibfnamefont{W.}~\bibnamefont{Losert}},
  \bibinfo{author}{\bibfnamefont{C.}~\bibnamefont{O'Hern}}, \bibnamefont{and}
  \bibinfo{author}{\bibfnamefont{R.}~\bibnamefont{Behringer}},
  \bibinfo{journal}{Phys. Rev. E} \textbf{\bibinfo{volume}{85}},
  \bibinfo{pages}{011305} (\bibinfo{year}{2012}).

\bibitem[{\citenamefont{Mason et~al.}(1999)\citenamefont{Mason, Levine,
  ErtaÅ, and Halsey}}]{Mason_99}
\bibinfo{author}{\bibfnamefont{T.}~\bibnamefont{Mason}},
  \bibinfo{author}{\bibfnamefont{A.}~\bibnamefont{Levine}},
  \bibinfo{author}{\bibfnamefont{D.}~\bibnamefont{ErtaÅ}}, \bibnamefont{and}
  \bibinfo{author}{\bibfnamefont{T.}~\bibnamefont{Halsey}},
  \bibinfo{journal}{Phys. Rev. E} \textbf{\bibinfo{volume}{60}},
  \bibinfo{pages}{R5044} (\bibinfo{year}{1999}).

\bibitem[{\citenamefont{Cundall and Strack}(1979)}]{cundall79}
\bibinfo{author}{\bibfnamefont{P.}~\bibnamefont{Cundall}} \bibnamefont{and}
  \bibinfo{author}{\bibfnamefont{O.}~\bibnamefont{Strack}},
  \bibinfo{journal}{G\'eotechnique} \textbf{\bibinfo{volume}{29}},
  \bibinfo{pages}{47} (\bibinfo{year}{1979}).

\end{thebibliography}
\end{document}